\documentclass[article, twocolumn,
   aps, pra,
  amsmath,amssymb,
  longbibliography,
  ]{revtex4-1}
\usepackage{graphicx,color}
\usepackage{amsmath}
\usepackage{natbib}
\usepackage{epsfig}
\begin{document}

\title{Freezing density scaling of fluid transport properties: Application to liquefied noble gases}

\author{S. A. Khrapak}\email{Sergey.Khrapak@gmx.de}
\affiliation{Joint Institute for High Temperatures, Russian Academy of Sciences, 125412 Moscow, Russia}
\author{A. G. Khrapak}
\affiliation{Joint Institute for High Temperatures, Russian Academy of Sciences, 125412 Moscow, Russia}

\begin{abstract}
A freezing density scaling of transport properties of the Lennard-Jones fluid is rationalized in terms of the Rosenfeld's excess entropy scaling and isomorph theory of Roskilde-simple systems. Then, it is demonstrated that the freezing density scaling operates reasonably well for viscosity and thermal conductivity coefficients of liquid argon, krypton, and xenon. Quasi-universality of the reduced transport coefficients at their minima and at freezing conditions is discussed. The magnitude of the thermal conductivity coefficient at the freezing point is shown to agree remarkably well with the prediction of the vibrational model of heat transfer in dense fluids. 
\end{abstract}

\date{\today}

\maketitle

\section{Introduction}

It has been recently demonstrated that properly reduced transport coefficients (self-diffusion, shear viscosity, and thermal
conductivity) of the Lennard-Jones (LJ) fluid along isotherms exhibit quasi-universal scaling on the density divided
by its value at the freezing point~\cite{KhrapakPRE04_2021}. 
This freezing density scaling (FDS) is a valuable result in view of the absence of a general theory of transport processes in fluids. In spite of a considerable progress achieved over the years~\cite{FrenkelBook,BarkerRMP1976,GrootBook,BalucaniBook,MarchBook,HansenBook},  
our understanding of fluid transport properties remains incomplete and fragmented. 

In the absence of general theories we often have to rely
on phenomenological approaches, semi-quantitative models, and scaling relationships. Among the most widely used are the Stokes-Einstein relation between the self-diffusion and the shear viscosity coefficients~\cite{BalucaniBook,ZwanzigJCP1983,Balucani1990,OhtoriPRE2015,
OhtoriPRE2017,
OhtoriJCP2018,CostigliolaJCP2019,
KhrapakMolPhys2019,OhtoriChemLett2020,
KhrapakPRE10_2021,KhrapakMolecules12_2021}
(correlations between the viscosity and thermal conductivity have been also discussed recently~\cite{KhrapakJETPLett2021}), excess entropy scaling of the reduced transport coefficients in fluids~\cite{RosenfeldPRA1977,RosenfeldJPCM1999,DzugutovNature1996,
DyreJCP2018,BellJPCB2019}, different variants of the freezing temperature scaling~\cite{RosenfeldPRE2000,RosenfeldJPCM2001,OhtaPoP2000,CostigliolaJCP2018,KhrapakAIPAdv2018}.  
These relationships operate in many simple (and sometimes not so simple) systems, but counterexamples when they are not applicable or questionable also exist (see e.g. Refs.~\cite{KrekelbergPRE03_2009,KrekelbergPRE12_2009,FominPRE2010,
DyreJCP2018,Ranieri2021,KhrapakJMolLiq2022} for some relevant examples). 

The FDS can be a useful and very convenient addition to the existing body of scaling relationships. 
In this paper we first provide a link between the FDS of transport properties of the LJ fluid proposed in Ref.~\cite{KhrapakPRE04_2021} and excess entropy scaling as well as  isomorph theory. Then we demonstrate that the quasi-universal FDS applies to the reduced viscosity and thermal conductivity coefficients of liquefied noble gases such as argon, krypton, and xenon. Some reference points for the reduced transport coefficients, such at their values at the minima and at freezing conditions are discussed. Explanation for quasi-universality of these values is provided. 
It is observed that the specific heat at constant volume also exhibits FDS, but only in the dense fluid regime, far above the critical point density.        

\section{Preliminaries}

The reference LJ potential is defined as 
\begin{equation}
\phi(r)=4\epsilon\left[\left(\frac{\sigma}{r}\right)^{12}-\left(\frac{\sigma}{r}\right)^{6}\right], 
\end{equation}
where  $\epsilon$ and $\sigma$ are the energy and length scales (or LJ units), respectively. The reduced density and temperature expressed in LJ units are $\rho^*=\rho\sigma^3$, $T^*=k_{\rm B}T/\epsilon$, where $k_{\rm B}$ is the Boltzmann constant. The LJ potential is one of the most popular and extensively studied models in condensed matter. It combines relative simplicity with adequate approximation of interatomic interactions in real substances and in particular can serve as a reasonable approximation for actual interactions in noble gases, liquids and solids. 

To make comparison between the transport properties of various real and model (LJ here) fluids feasible, one should use a rational normalization for the transport coefficients. Particularly useful macroscopically reduced units for the viscosity and thermal conductivity are       
\begin{equation}\label{Rosenfeld}
\eta_{\rm R}  =  \eta \frac{\rho^{-2/3}}{m v_{\rm T}}, \quad\quad \lambda_{\rm R}=\lambda\frac{\rho^{-2/3}}{v_{\rm T}},
\end{equation}
where $\eta$ and $\lambda$ are the shear viscosity and thermal conductivity coefficients, $v_{\rm T}=\sqrt{k_{\rm B}T/m}$ is the thermal velocity, and $m$ is the atomic mass.
This normalization is essential in the Rosenfeld's excess entropy scaling approach~\cite{RosenfeldPRA1977,RosenfeldJPCM1999}, therefore the subscript ${\rm R}$ is often used (although in fact similar normalization is traceable to earlier works). 

\section{Origin of FDS}

Originally, the FDS of transport coefficients in the LJ fluid has been proposed as merely an empirical observation~\cite{KhrapakPRE04_2021}. The premise to look into possible FDS was based on various forms of freezing temperature scaling discussed in the literature~\cite{RosenfeldPRE2000,RosenfeldJPCM2001,CostigliolaJCP2018,Kaptay2005}. An important  example is related to the screened Coulomb (Yukawa) interaction potential, widely investigated in the context of plasma-related and colloidal interactions~\cite{FortovUFN,FortovPR,IvlevBook,ChaudhuriSM2011}.
Here the static and dynamical properties depend on the dimensionless coupling parameter $\Gamma=q^2\rho^{1/3}/k_{\rm B}T$, where $q$ is the particle charge. Various thermodynamic~\cite{RosenfeldPRE2000,ToliasPRE2014,KhrapakPRE02_2015,KhrapakPRE03_2015,
KhrapakJCP2015,CastelloPRE2021,CastelloJCP2021}, collective~\cite{DonkoJPCM2008}, as well as transport properties~\cite{OhtaPoP2000,SaigoPoP2002,VaulinaPRE2002,DonkoPRE2004,
KhrapakPoP2012,KhrapakAIPAdv2018} are to a good accuracy determined by the ratio $\Gamma/\Gamma_{\rm fr}$ alone, where $\Gamma_{\rm fr}$ is the coupling parameter at the freezing point.  Obviously, a universal scaling with $\Gamma/\Gamma_{\rm fr}$ implies not only a freezing temperature scaling (universal dependence on $T/T_{\rm fr}$), but also a FDS (universal dependence on $\rho/\rho_{\rm fr}$). This was the main motivation to examine applicability of a FDS to the LJ fluid~\cite{KhrapakPRE04_2021}.          

More recently, FDS has been discussed in the context of excess entropy scaling and the theory of isomorphs~\cite{KhrapakJPCL2022}. Let us elaborate on this further here. 
We first remind that the reduced excess entropy is defined as $s_{\rm ex} = (S - S_{\rm id})/N$, where $S$ is the system entropy, $S_{\rm id}$ is the entropy of the
ideal gas at the same temperature and density, and $N$ is the
number of particles. Sometimes $s_{\rm ex}$ is measured in units of $k_{\rm B}$. For convenience, we introduce a new excess entropy variable $s^*=s_{\rm ex}/k_{\rm B}$, so that $s^*$ becomes dimensionless. It is negative, because interatomic interactions increase ordering compared to a fully disordered ideal gas. 
In the fluid state $s^*$ can range between zero (ideal non-interacting gas) and $\simeq -4$ (at the freezing point, see below).

 Isomorph theory predicts that many liquids exhibit an approximate
``hidden'' scale invariance that implies the existence of lines
in the thermodynamic phase diagram, so-called isomorphs,
along which structure and dynamics in properly reduced units
are invariant to a good approximation~\cite{GnanJCP2009,SchroderJCP2014,DyreJPCB2014}. Excess entropy is
constant along isomorphs and this relates excess entropy scaling to the isomorphs theory~\cite{DyreJCP2018}. It turns out that the lines corresponding to constant ratios $\rho/\rho_{\rm fr}$ correspond to quasi-constant values of the excess entropy $s^*$. To demonstrate this we apply an equation of state (EoS) for the LJ fluid proposed by Thol {\it et al}.~\cite{Thol2016} and calculate the lines of constant excess entropy. We choose three values of the reduced excess entropy, $s^*=-1$, $-2$, and $-3$. 

The line $s^*=-1$ corresponds roughly to the dynamical crossover between the gas-like and liquid-like behaviours (so-called ``Frenkel line on the phase diagram'')~\cite{BrazhkinPRE2012,BrazhkinUFN2012,BellJCP2020,BellJPCL2021,
KhrapakJCP2022}. A particular value $s^*\simeq -1$ emerges from the intersection of gas-like and liquid-like asymptotes of the Stokes-Einstein product $D\eta/\rho^{1/3}k_{\rm B}T$, where $D$ is the self-diffusion coefficient~\cite{KhrapakJCP2022,KhrapakPRE10_2021}. Another definition, based on the location of the minimum of the reduced shear viscosity coefficient as a function of excess entropy leads to $s^*=-2/3$~\cite{BellJCP2020}. Other definitions may result in somewhat different values. This is not crucial, however, because a crossover rather than a sharp transition takes place.

The line $s^*=-2$ marks the onset of the dense fluid regime. For $s^*\lesssim -2$ the Stokes-Einstein relation without the hydrodynamic radius is satisfied to a very good accuracy in LJ and some other simple fluids~\cite{KhrapakPRE10_2021}. A solid-like cell theory approach for the excess entropy becomes appropriate for different purely repulsive interactions (although not for the LJ system) at $s^*\lesssim-2$~\cite{KhrapakJCP2021}.

The value $s^*=-3$ does not apparently have any special significance and is plotted for completeness. 

\begin{figure}
\includegraphics[width=7.5cm]{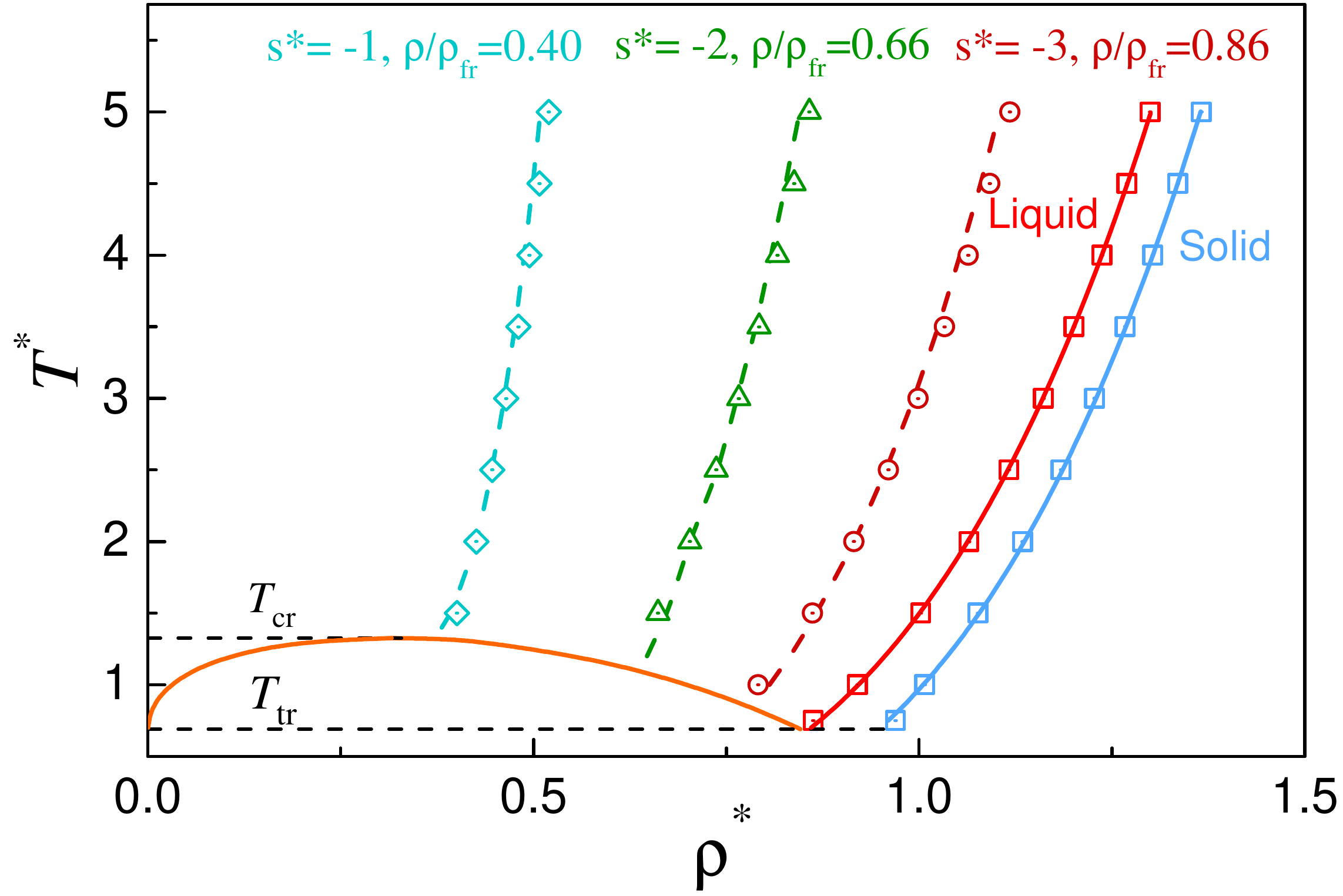}
\caption{(Color online) Phase diagram of the Lennard-Jones system. The squares correspond to the fluid-solid coexistence boundaries as tabulated in Ref.~\cite{SousaJCP2012}; the corresponding curves are simple fits to the functional form $T^*={\mathcal A}(\rho^*)^4+{\mathcal B}(\rho^*)^2$ (for the freezing curve we have used ${\mathcal A}=2.111$ and ${\mathcal B}=0.615$).  The liquid-vapour boundary is plotted using the formulas provided in Ref.~\cite{HeyesJCP2019}. 
Three dashed curves correspond to constant values of the excess entropy, $s^*=-1$, $s^*=-2$, and $s^*=-3$ (from left to right), as evaluated from the LJ equation of state of Thol {\it et al}.~\cite{Thol2016}. Symbols correspond to constant values of reduced density, $\rho/\rho_{\rm fr}=0.40$, $\rho/\rho_{\rm fr}=0.66$, $\rho/\rho_{\rm fr}=0.86$ (also from left to right). This demonstrates that curves characterized by a fixed ratio $\rho/\rho_{\rm fr}$ are also characterized by a quasi-constant value of $s^*$ and thus are approximate isomorphs.}
\label{Fig1}
\end{figure}

The phase diagram of the LJ system is shown in Fig.~\ref{Fig1}. The data for the phase coexistence boundaries are taken from Refs.~\cite{SousaJCP2012,HeyesJCP2019}. The fits for the freezing and melting curves are similar to the wide-temperature-range   
equation from Ref.~\cite{KhrapakJCP2011_2}, but deliver somewhat better accuracy in the regime $0.75<T^*<5$. The three lines of constant excess entropy are shown by dashed curves. The superimposed symbols correspond to constant ratios of $\rho/\rho_{\rm fr}$, equal to $0.40$ (for $s^*=-1$), $0.66$ (for $s^*=-2$), and $0.86$ (for $s^*=-3$), as obtained from the data tabulated in Ref.~\cite{SousaJCP2012}. We observe that the constancy of $\rho/\rho_{\rm fr}$ implies approximate constancy of $s^*$ and, therefore, these lines are approximate isomorphs. 

We should point out here that by definition of Roskilde-simple systems they are characterized by strong correlations between the virial and potential energy. A practical requirement usually employed is that the Pearson correlation coefficient $R$ exceeds $\simeq 0.9$~\cite{GnanJCP2009,DyreJPCB2014}. For the LJ fluid this happens at near-freezing densities and at high temperatures~\cite{BaileyJCP2013,BellJPCB2019}. On approaching the gas-liquid coexistence boundary and the critical point, $R$ decreases considerably. However, even in this regime excellent correlations between the transport properties and excess entropy persist~\cite{BellJPCB2019}. Apparently, the applicability range of excess entropy scaling is wider than that guaranteed by the condition $R\gtrsim 0.9$ in this case. Consequently, it might be more appropriate to consider FDS as a manifestation of the excess entropy scaling.   

\begin{figure}
\includegraphics[width=7.5cm]{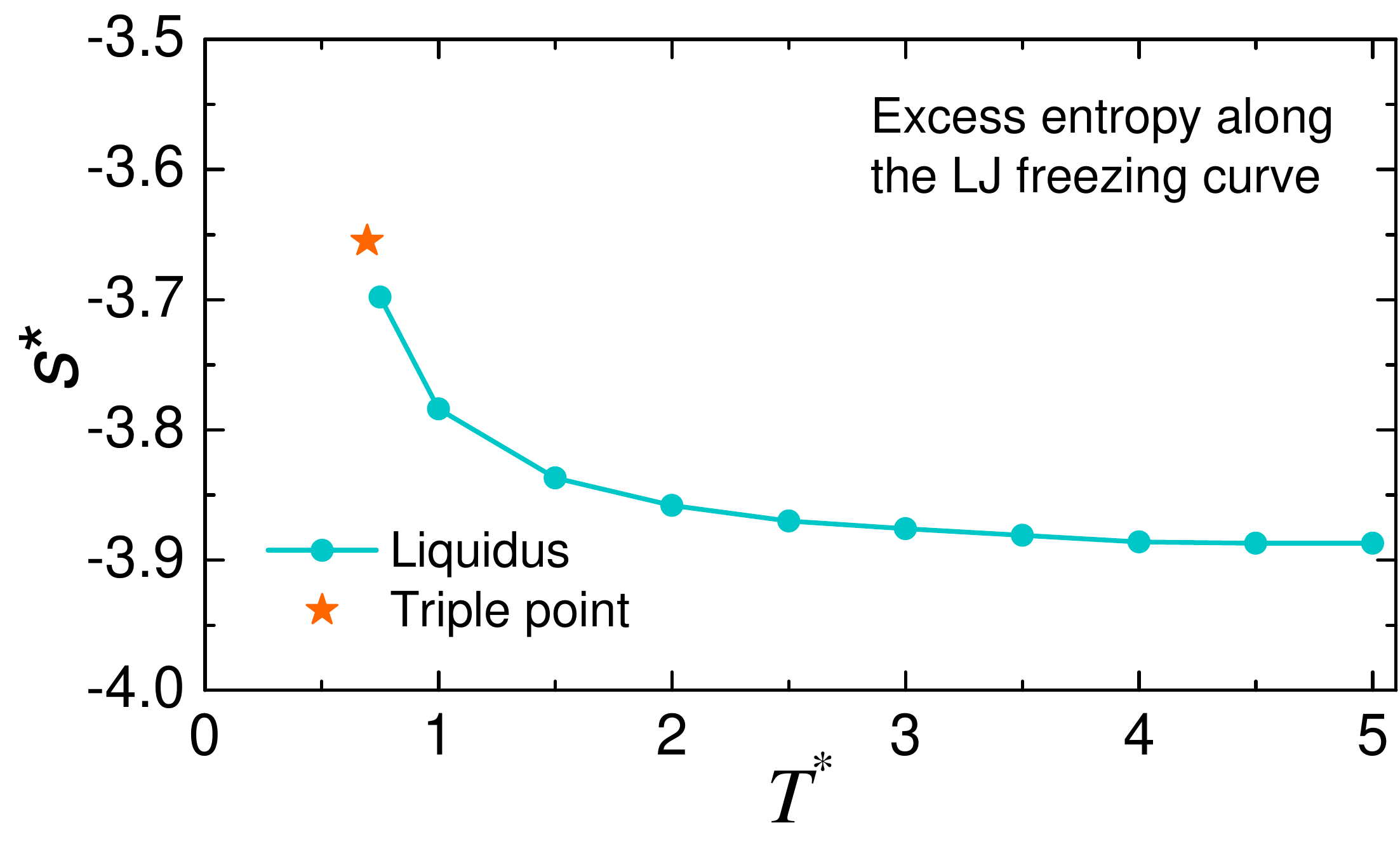}
\caption{(Color online) Reduced excess entropy, $s^*$, at the freezing line of the Lennard-Jones fluid as a function of the reduced temperature $T^*$. }
\label{Fig2}
\end{figure}

The excess entropy along the freezing curve of the LJ fluid decreases with temperature from $s^*\simeq -3.65$ at the triple point and saturates at $\simeq -3.9$ at higher temperatures. This is illustrated in Fig.~\ref{Fig2}, where $s^*$ is plotted as a function of $T^*$. To calculate this dependence we have used the liquidus data tabulated in Ref.~\cite{SousaJCP2012} along with Thol {\it et al.} EoS~\cite{Thol2016}.

\begin{figure}
\includegraphics[width=7.5cm]{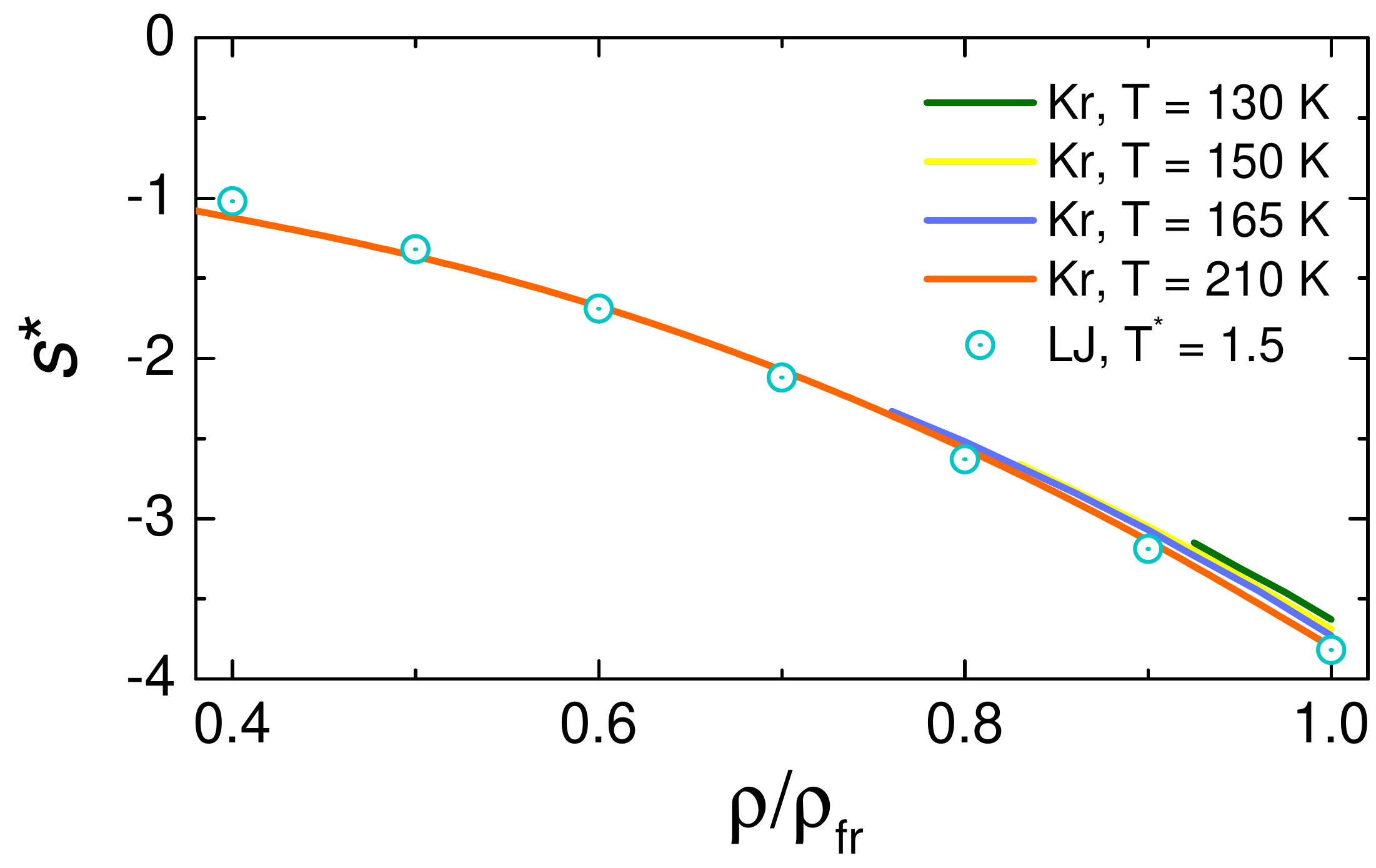}
\caption{(Color online) Reduced excess entropy $s^*$ versus the reduced density $\rho/\rho_{\rm fr}$ in liquefied krypton. Solid curves correspond to isotherms $T=130$, 150, 165, and 210 K (see the legend). Entropies are calculated using the EoS from Ref.~\cite{Lemmon2006}. Symbols correspond to a reference LJ isotherm $T^*=1.5$ as calculated from the EoS of Ref.~\cite{Thol2016}.}
\label{Fig8}
\end{figure}

Finally, Figure~\ref{Fig8} illustrates the freezing density scaling of the excess entropy in liquid krypton at three subcritical ($T=130, 150$ and 165 K) isotherms and one near-critical isotherm $T=210$ K. The excess entropy has been evaluated from the EoS by Lemmon and Span~\cite{Lemmon2006}. The figure demonstrates that the dependence of $s^*$ on the reduced density $\rho/\rho_{\rm fr}$ is approximately quasi-universal for both subcritical and supercritical temperatures. Thus excess entropy scaling of transport coefficients implies the freezing density scaling and vice versa. Moreover, it is observed that the dependence of $s^*$ on $\rho/\rho_{\rm fr}$ for krypton is very similar to that in the LJ fluid. This is a strong indication that the LJ FDS should operate in liquefied noble gases. This conjecture shall be now verified in detail.        

\section{Application to liquid noble gases}

We have obtained the reduced viscosity and thermal conductivity coefficients of argon, krypton and xenon liquids using the data provided in Institute of Standards and Technology (NIST) Reference Fluid Thermodynamic and Transport Properties Database (REFPROP 10)~\cite{Refprop}. For each fluid three isotherms have been selected, $T=0.75 T_{\rm c}$,  $T=1,25 T_{\rm c}$, and  $T=1.50 T_{\rm c}$, where $T_{\rm c}$ is the critical temperature. We did not included helium liquid into consideration, because in this case quantum effects can interfere~\cite{HansenPR1969}. For neon an inconsistency between the data from \cite{Refprop} and \cite{NIST} has been recently pointed out~\cite{KhrapakPoF2022}. This issue should be addressed, before detailed analysis can be performed. 

\begin{table}
\caption{\label{Tab1}{Individual models for viscosity, thermal conductivity, equation of state, and melting curve of argon, krypton, and xenon, as implemented in REFPROP 10.}}
\begin{ruledtabular}
\begin{tabular}{crrr}
Liquid & argon & krypton & xenon   \\ \hline
$\eta$ & Ref.~\cite{LemmonIJT2004} & Ref.~\cite{Huber2018} & Ref.~\cite{Huber2018}    \\
$\lambda$ & Ref.~\cite{LemmonIJT2004} &  Ref.~\cite{Huber2018} & Ref.~\cite{Huber2018} \\
EoS & Ref.~\cite{Tegeler1999} & Ref.~\cite{Lemmon2006}  & Ref.~\cite{Lemmon2006}   \\
Melting curve &  Ref.~\cite{Tegeler1999} & Ref.~\cite{Michels1962} & Ref.~\cite{Michels1962}  \\
\end{tabular}
\end{ruledtabular}
\end{table}

The individual models for viscosity, thermal conductivity, equation of state, and melting curve for each species implemented in REFPROP 10 can be found in project documentation~\cite{Refprop} and in Refs~\cite{LemmonIJT2004,Lemmon2006,Huber2018,Tegeler1999,Michels1962}.
For convenience we have summarized this information in Table~\ref{Tab1}.
The quoted uncertainties in transport coefficients are up to $\sim 5\%$ for argon and xenon~\cite{Refprop}. The documentation of REFPROP 10 states about the model for krypton: ``Uncertainty of viscosity in the liquid phase is 30$\%$, data unavailable. Uncertainty of viscosity in the gas phase at atmospheric pressure is 3$\%$. Uncertainty of thermal conductivity is 4$\%$ at pressures to 50 MPa''~\cite{Refprop}.

\begin{figure}
\includegraphics[width=7.5cm]{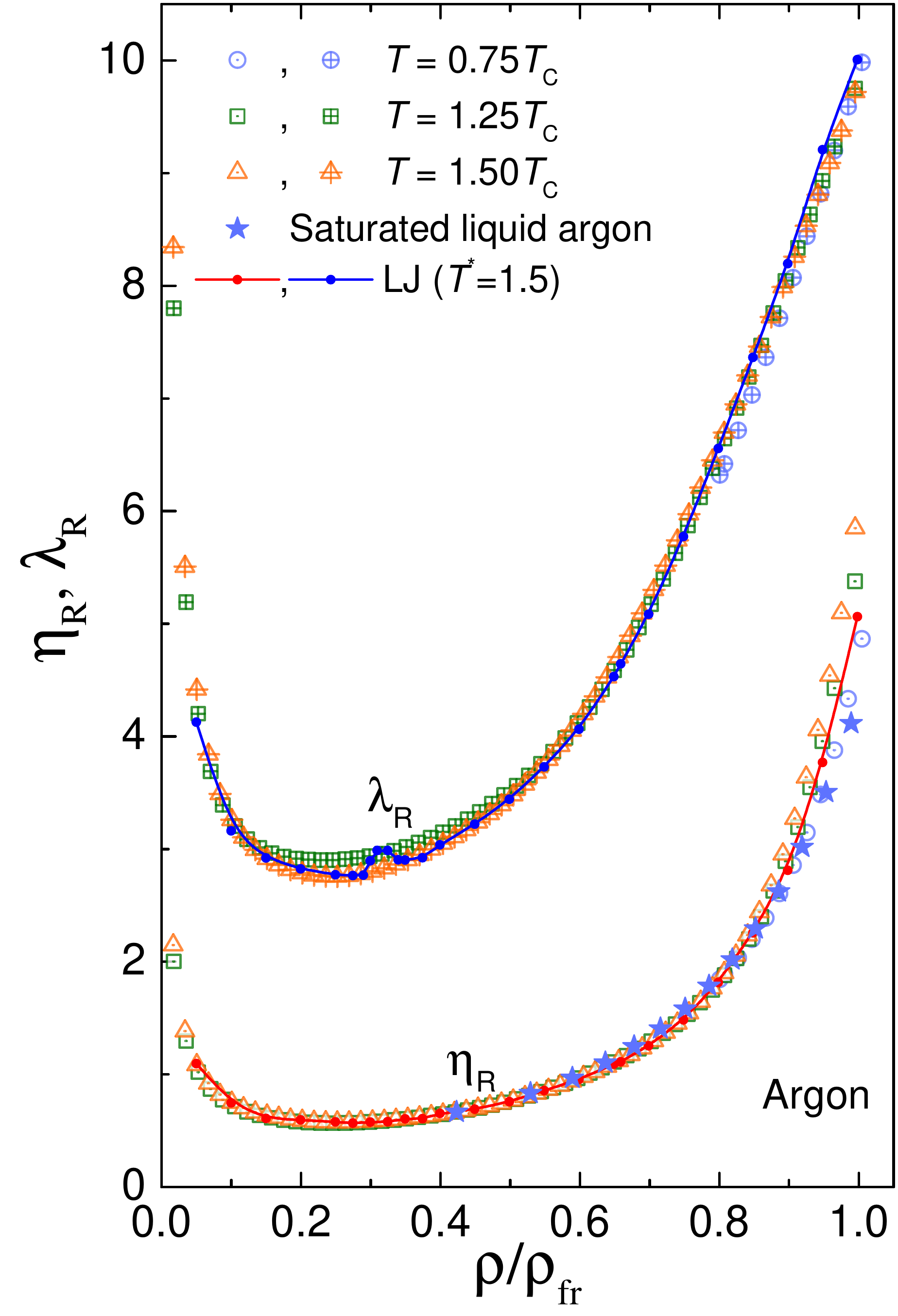}
\caption{(Color online) Reduced viscosity and thermal conductivity coefficients of argon liquid, $\eta_{\rm R}$ and $\lambda_{\rm R}$, along the isotherms versus the reduced density $\rho/\rho_{\rm fr}$. Data for three isotherms, $T=0.75 T_{\rm c}$,  $T=1.25 T_{\rm c}$, and  $T=1.50 T_{\rm c}$, are taken from REFPROP 10~\cite{Refprop}. Filled stars are the recommended viscosity data for saturated liquid argon tabulated in Ref.~\cite{HanleyJPCRD1974}. Small circles connected by solid curves correspond to the reference data for the LJ fluid isotherm $T^*=1.5$ from Refs.~\cite{BaidakovJCP2012,BaidakovJCP2014}.  }
\label{Fig3}
\end{figure}

\begin{figure}
\includegraphics[width=7.5cm]{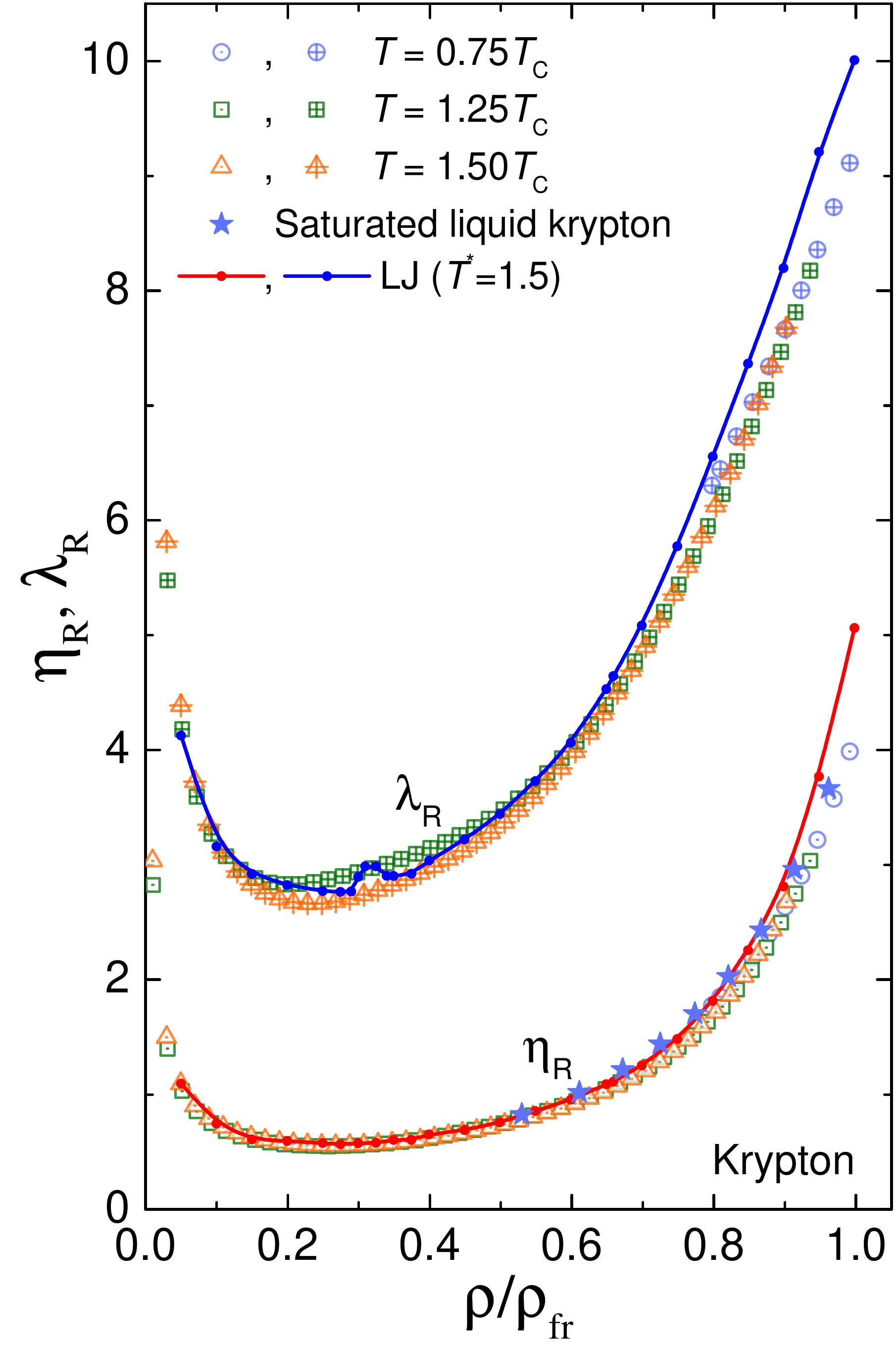}
\caption{(Color online) Same as in Fig.~\ref{Fig3}, but for Krypton liquid. }
\label{Fig4}
\end{figure}

\begin{figure}
\includegraphics[width=7.5cm]{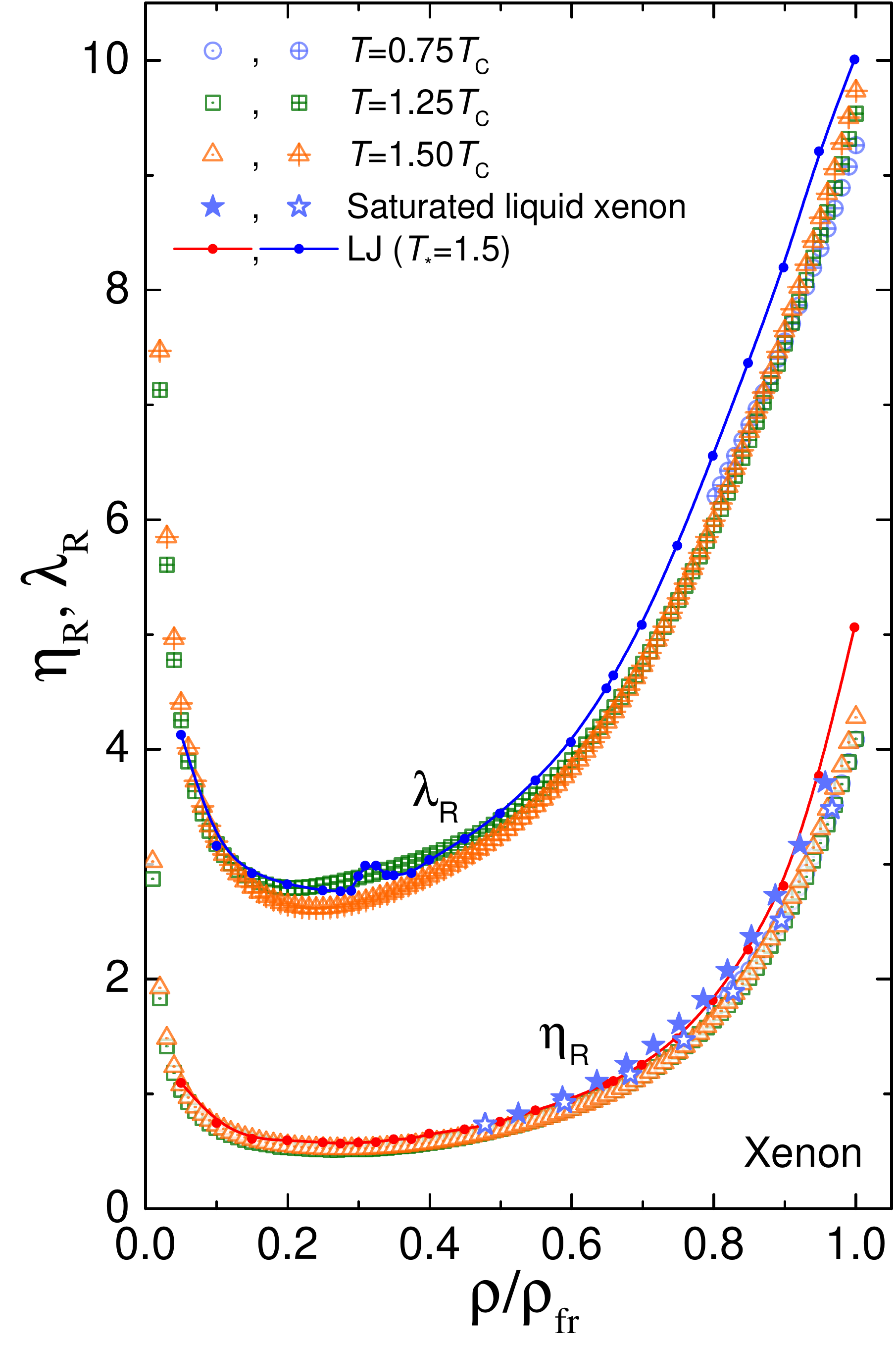}
\caption{(Color online) Same as in Fig.~\ref{Fig3}, but for Xenon liquid. Open stars correspond to viscosity values of saturated liquid xenon, calculated a new reference correlation scheme from Ref.~\cite{Velliadou2021}.}
\label{Fig5}
\end{figure}

Reduced viscosity ($\eta_{\rm R}$) and thermal conductivity ($\lambda_{\rm R}$) coefficients along selected isotherms, as obtained from REFPROP 10, are plotted versus the reduced density $\rho/\rho_{\rm fr}$ for argon, krypton and xenon in Figs.~\ref{Fig3} -- \ref{Fig5}. We use symbols for REFPROP data just for better readability. It should be reminded that REFPROP does not contain experimental data, but rather models based on experimental data. The data for $\lambda_{\rm R}$ are always above those for $\eta_{\rm R}$. Recommended reference viscosity data~\cite{HanleyJPCRD1974} for saturated liquid argon, krypton, and xenon are also plotted in Figs.~\ref{Fig3} -- \ref{Fig5}. Small circles connected by solid curves are the reference FDS for the LJ fluid. A comprehensive data collection for the transport coefficients of the LJ fluid can be found for instance in Ref.~\cite{BellJPCB2019}. Here we use a single representative dataset corresponding to the LJ isotherm $T^*=1.5$, as evaluated in Refs.~\cite{BaidakovJCP2012,BaidakovJCP2014}. 
An important property of the FDS is that the length and energy scales of the LJ potential are not involved, and thus the quality of description does not depend on their correct choice. 

The first important observation in Figs.~\ref{Fig3} -- \ref{Fig5} is that the FDS holds for each of the considered liquids. Reduced transport coefficients along different isotherms do coincide when plotted against $\rho/\rho_{\rm fr}$. The only case when some scattering is observed is related to the viscosity coefficient of argon in the vicinity of the freezing point (Fig.~\ref{Fig3}). The scattering is, however, not too large and $\eta_{\rm R}$ remains in a relatively narrow range $\eta_{\rm R}\simeq 5.3\pm 0.5$ at near-freezing densities. Actually, the freezing line is approximately an isomorph (see e.g. Fig.~\ref{Fig2} for the LJ fluid and Ref.~\cite{SinghJCP2021} for simulations of solid-liquid coexistence in noble gases) and hence reduced transport coefficients are not expected to vary much along this line~\cite{CostigliolaPCCP2016,PedersenNatCom2016}. Some variations have to be nevertheless expected. The magnitude of these variations can be anticipated as follows. According to the modified excess entropy scaling proposed in Ref.~\cite{BellJPCB2019}, the excess component of the viscosity (dominating at near-freezing densities) changes with excess entropy from $\eta_{\rm R}\simeq 4.4$ at $s^*=-3.65$ (near the triple point) to  $\eta_{\rm R}\simeq 5.7$ at $s^*=-3.9$ (at higher temperatures). This correlates with the picture presented in Fig.~\ref{Fig3}, illustrating viscosity coefficient of liquefied argon. For krypton and xenon, however, no clear dependence of $\eta_{\rm R}$ on $T^*$ at near-freezing conditions is observed.        

\begin{figure}
\includegraphics[width=8cm]{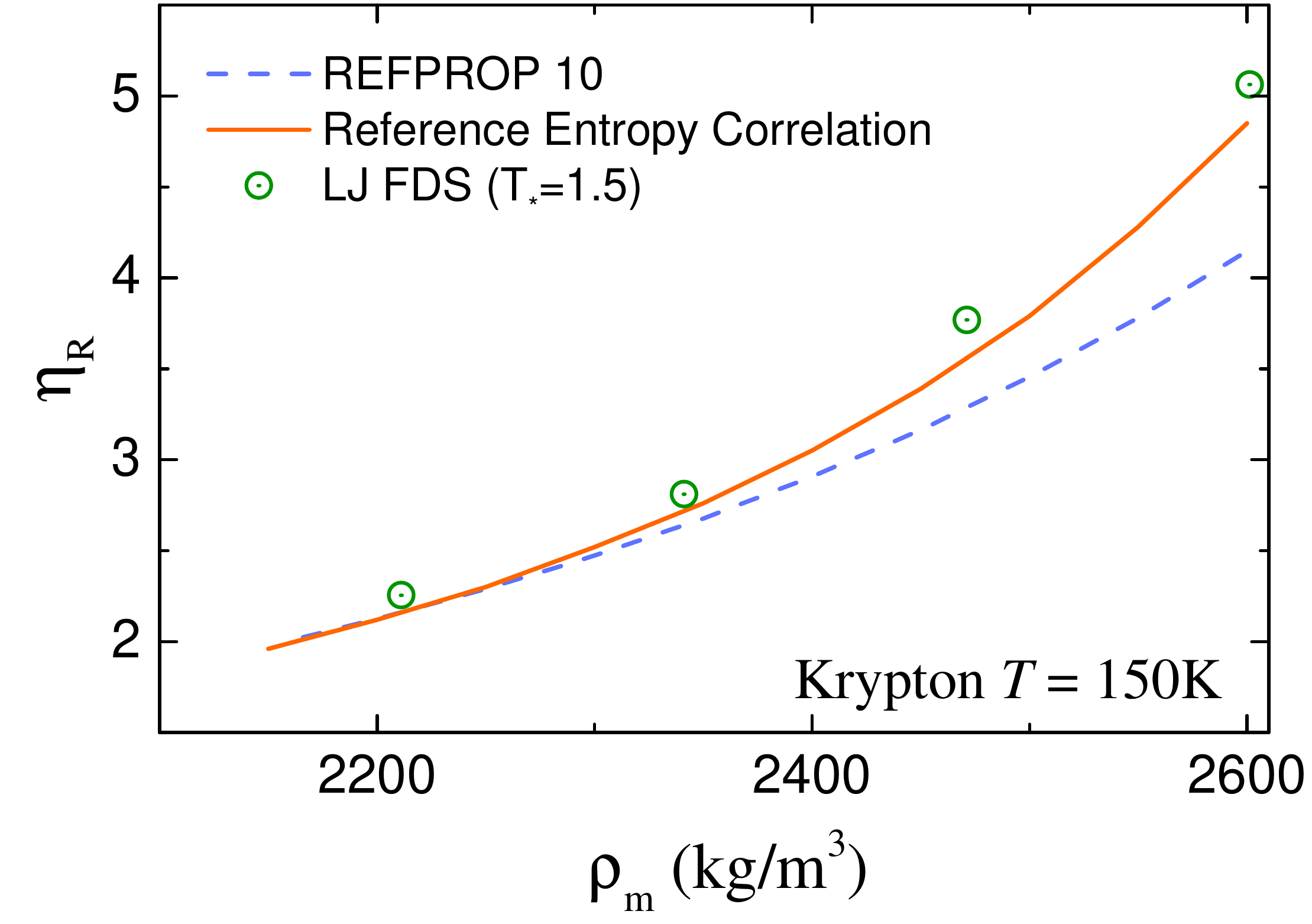}
\caption{(Color online) Reduced viscosity of liquid krypton at $T=150$ K as a function of mass density $\rho_m$. Dashed curve corresponds to REFPROP 10, solid curve is calculated using a reference correlation from the excess entropy scaling~\cite{Polychroniadou2021}. Symbols correspond to LJ FDS using reference data for the LJ fluid isotherm $T^*=1.5$~\cite{BaidakovJCP2012}.}
\label{Fig7}
\end{figure}

The second observation is related to the quality of the reference LJ scaling. We observe that for argon the REFPROP data are very well consistent with the LJ scaling (see Fig.~\ref{Fig3}). In cases of krypton (Fig.~\ref{Fig4}) and xenon (Fig.~\ref{Fig5}) LJ scaling overestimates the REFPROP transport data systematically in the vicinity of the freezing transition. The relative deviations are not too large (in particular taking into account quoted uncertainty for krypton). On the other hand, the recommended data on the viscosity of saturated liquids tabulated in Ref.~\cite{HanleyJPCRD1974} seem to be somewhat better correlated with the reference LJ fit, compared to models used in REFPROP. Results from the new correlation scheme for the viscosity of xenon~\cite{Velliadou2021} are shown in Fig.~\ref{Fig5} by open stars. These are in good agreement with the model already implemented in REFPROP 10, but deviate from the tabulated recommended values from Ref.~\cite{HanleyJPCRD1974} as well as LJ FDS. On the other hand, in a recent paper ~\cite{Polychroniadou2021} it has been demonstrated that the dependence of the reduced excess component of the viscosity of krypton on the excess entropy is very well correlated with that of the LJ fluid. This suggests that an improved model of the viscosity of krypton can be more consistent with the LJ scaling. This is illustrated in Fig.~\ref{Fig7}. Indeed, the new correlation that uses the excess entropy scaling~\cite{Polychroniadou2021} (solid curve) is closer to the LJ FDS (circles) than the results from REFPROP 10 (dashed curve). This example also demonstrates that significant deviations from the freezing density scaling may be considered as a signal to scrutinize the consistency of this data.    

The LJ data for the thermal conductivity coefficient along $T^*=1.5$ isotherm exhibit a critical enhancement at a near critical density. This is not seen in REFPROP data, apparently because the corresponding isotherms are further away from the critical one. Nevertheless, we observe that the reduced thermal conductivity coefficient $\lambda_{\rm R}$ at $T=1.25T_{\rm c}$ is systematically larger than that at  $T=1.50T_{\rm c}$ at near-critical densities, which can be another manifestation of the critical enhancement. For the viscosity coefficient critical enhancement can normally be ignored~\cite{BellJPCB2019}.   

Let us now focus on the shape of the curves. They exhibit clear minima when the reduced density increases from zero to unity. The origin of the minima in the reduced viscosity and thermal conductivity coefficients is the crossover between the gas-like and liquid-like mechanisms of the momentum and energy transfer. Recently, it has been suggested that the {{\it kinematic viscosity} and {\it thermal diffusivity}} of liquids and supercritical fluids have lower bounds determined by fundamental physical constants~\cite{TrachenkoSciAdv2020,TrachenkoPRB2021,TrachenkoPhysToday2021}. There exist, however, purely classical arguments, which suggest that macroscopically reduced viscosity and thermal conductivity coefficients can be expected to reach quasi-universal values at their respective minima~\cite{KhrapakPoF2022}. For several model and real monatomic fluids considered in that work it was observed that $\eta_{\rm R}^{\rm min} = 0.6\pm 0.1$ and $\lambda_{\rm R}\simeq 2.8 \pm 0.2$. The only exception identified was the one-component plasma fluid, where the minimal values are considerably lower due to extremely soft and long-ranged character of the Coulomb interaction potential. The data presented in Figs.~\ref{Fig3}--\ref{Fig5} are consistent with these earlier observations. The minima are located in a relatively narrow density range $\rho/\rho_{\rm fr}\simeq 0.2-0.3$ and their magnitude are $\eta_{\rm R}^{\rm min}\simeq 0.57$ and $\lambda_{\rm R}^{\rm min}\simeq 2.8-2.9$ for argon, $\eta_{\rm R}^{\rm min}\simeq 0.55$ and $\lambda_{\rm R}^{\rm min}\simeq 2.7-2.8$ for krypton, and $\eta_{\rm R}^{\rm min}\simeq 0.51$ and $\lambda_{\rm R}^{\rm min}\simeq 2.6-2.8$ for xenon. For the thermal conductivity coefficient the lower range limit corresponds to the isotherm $T=1.50 T_{\rm c}$, while the upper limit to the isotherm $T=1.25 T_{\rm c}$. 


\begin{figure}
\includegraphics[width=8cm]{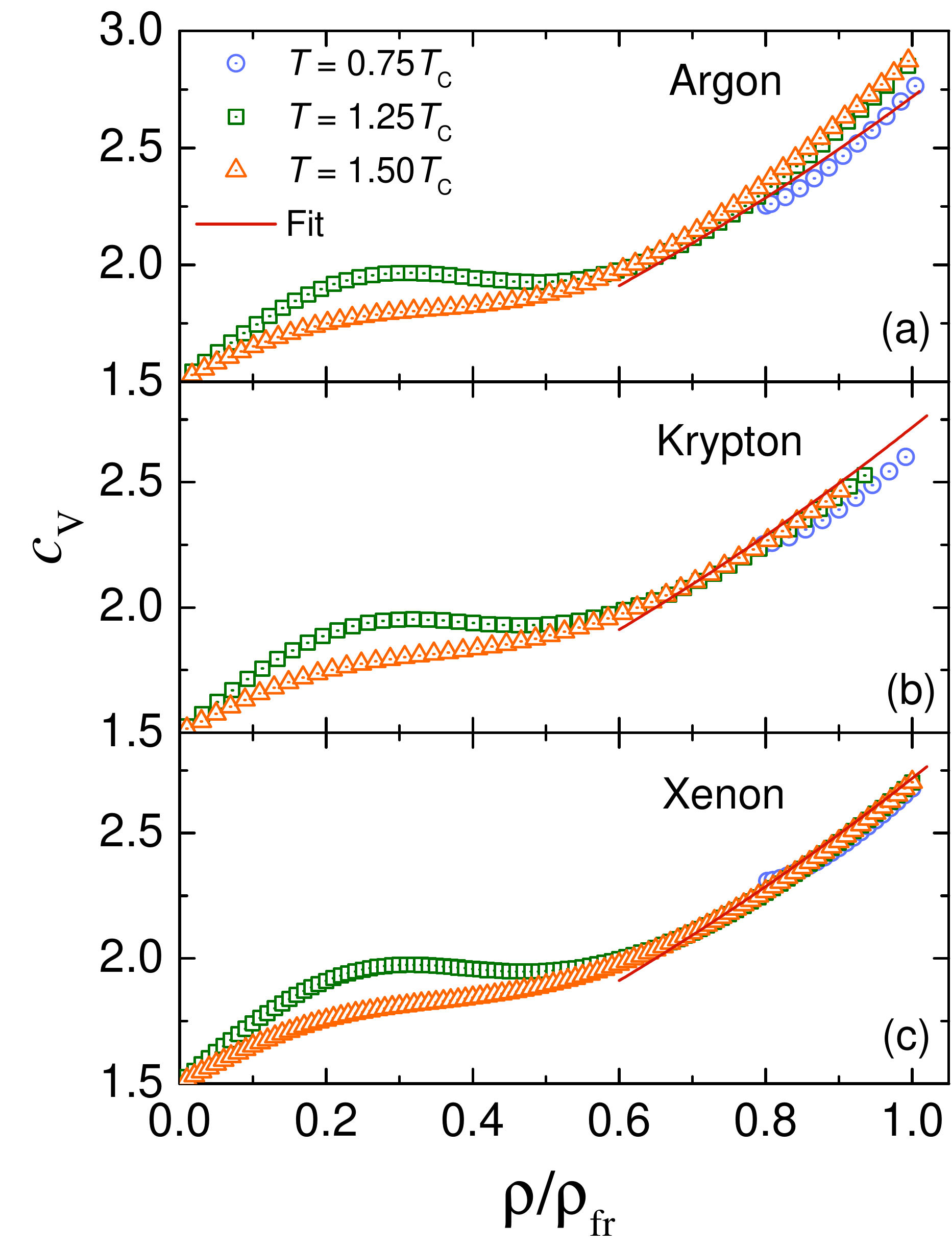}
\caption{(Color online) Specific heat at constant volume $c_{\rm v}$ vs the reduced density $\rho/\rho_{\rm fr}$ for argon (a), krypton (b), and xenon (c). Data for three isotherms, $T=0.75 T_{\rm c}$,  $T=1.25 T_{\rm c}$, and  $T=1.50 T_{\rm c}$, are taken from REFPROP 10~\cite{Refprop}. The solid curves correspond to a simple empirical quasi-universal fit of Eq.~(\ref{cv}).}
\label{Fig6}
\end{figure}

Figure~\ref{Fig6} demonstrates the dependence of the specific heat at constant volume $c_{\rm v}$ on the reduced density. Here the FDS cannot be universal in a wide density range, because $c_{\rm v}$ is greatly affected in the vicinity of the critical point. Nevertheless, the curves for different liquids at the same ratio $T/T_{\rm c}$ are close. Moreover, in the dense fluid regime with $\rho/\rho_{\rm fr}\gtrsim 0.65$ ($s^*\lesssim -2$), all the data sets can be relatively well described by a simple empirical fit 
\begin{equation}\label{cv}
c_{\rm v}\simeq \exp\left[\left(\rho/\rho_{\rm fr}\right)^{0.85}\right].
\end{equation} 
This fit is particularly appropriate for xenon. For argon and krypton deviations are observable, but they do not exceed $\sim 5\%$. At the freezing point $c_{\rm v}\sim 2.7$ (and slightly higher for argon), which is somewhat below the Dulong-Petit value of $c_{\rm v}=3$ for a high-temperature classical solids. The observed quasi-universality can serve as a useful reference point to test the accuracy of various models of liquid thermodynamics, such as for instance phonon- and collective modes-based approaches discussed recently~\cite{BolmatovSciRep2012,BolmatovAP2015,TrachenkoRPP2015,
KryuchkovPRL2020}.     

From the isomorph theory and excess entropy scaling perspectives we should expect that the values of the reduced transport coefficients are quasi-universal along the freezing line of each substance. This is because freezing line is approximately an isomorph and excess entropy remains approximately constant. Significant changes from one substance to another are also not expected. Freezing of simple monatomic liquids occurs at about the same value of excess entropy, $s^*\simeq -4$~\cite{RosenfeldPRE2000}. For example, the LJ fluid freezes in a narrow range between $s^*\simeq -3.65$ and $s^*\simeq -3.9$, as shown in Fig.~\ref{Fig2}. Historically, the approximate constancy of the reduced viscosity and thermal conductivity coefficients at the freezing point can be traced back to the works of Andrade~\cite{Andrade1931,Andrade1934,Andrade1952}.

For the reduced viscosity coefficient, the FDS scaling in the LJ fluid predicts $\eta_{\rm R}\simeq 5.0$ at freezing. This is in satisfactory agreement with the data for argon, krypton, and xenon from REFPROP 10, taking into account their uncertainties. This is also comparable with the freezing point viscosities reported for other fluids: For example $\eta_{\rm R}\simeq 5.2$ and $\eta_{\rm R}\simeq 5.8$ in high-temperature LJ fluid and liquid argon~\cite{CostigliolaJCP2018,PedersenNatCom2016}; $\eta_{\rm R}\simeq 4.8$ for Coulomb and screened Coulomb (Yukawa) fluids~\cite{KhrapakAIPAdv2018}; $\eta_{\rm R}$ between $\simeq 5$ and $\simeq 6$ for many liquid metals at their respective freezing points~\cite{KhrapakAIPAdv2018,MarchBook}. A very few predictive model, which allow to relate the exact value of the reduced viscosity at the freezing point to the properties of interaction potential or thermodynamic parameters, exist (see e.g. Ref.~\cite{BellJCED2020} for a recent example).   
      
A predictive model for the thermal conductivity coefficient of simple monatomic liquids has been proposed recently~\cite{KhrapakPRE01_2021}. In this vibrational model of thermal conduction, the thermal conductivity coefficient can be expressed using fluid density, specific heat $c_{\rm v}$, and instantaneous (infinite frequency) longitudinal and transverse sound velocities $c_l$ and $c_t$:
\begin{equation}\label{Cahill1}
\lambda \simeq \frac{1}{4}\left(\frac{3}{4\pi}\right)^{1/3} c_{\rm v} \rho^{2/3}\left(c_l+2c_t\right). 
\end{equation}
The longitudinal and transverse sound velocities appear as a result of averaging over one longitudinal and two transverse collective modes, assuming that they exhibit acoustic dispersion. Equation (\ref{Cahill1}) resembles the one from the so-called minimal thermal conductivity model by Cahill and Pohl~\cite{Cahill1989,Cahill1992}, which is in good agreement with the measured thermal conductivities of many
amorphous inorganic solids, highly disordered crystals, and amorphous macromolecules~\cite{XiePRB2017}. Equation (\ref{Cahill1}) contains no free parameters and has been shown to describe relatively well the numerical data for LJ and Yukawa fluids~\cite{KhrapakPRE01_2021,KhrapakPoP08_2021}.

In reduced units Eq.~(\ref{Cahill1}) becomes
\begin{equation}\label{Cahill2}
\lambda_{\rm R}\simeq 0.155c_{\rm v}\frac{c_l+2c_t}{v_{\rm T}}. 
\end{equation}
The ratios $c_l/v_{\rm T}$ and $c_t/v_{\rm T}$ at the liquid-solid coexistence of the LJ fluid have been recently evaluated~\cite{KhrapakMolecules2020}. The calculated ratios are practically constant along the melting and freezing curves. The characteristic numerical values are $c_l/v_{\rm T}\simeq 11.5$ and $c_t/v_{\rm T}\simeq 6$. Substituting these values into Eq.~(\ref{Cahill2}) together with $c_{\rm v}\simeq 2.7$ at freezing of the LJ fluid and liquid noble gases, we arrive at $\lambda_{\rm R}\simeq 9.8$. This estimate is in rather good agreement with the data plotted in Figs.~\ref{Fig3} -- \ref{Fig5}, providing yet another confirmation of the adequacy of the vibrational model of thermal conductivity in dense fluids~\cite{KhrapakPRE01_2021}.       

\section{Conclusion}

In this paper we have first demonstrated that the freezing density scaling of transport coefficients in the LJ fluid is closely related to their excess entropy scaling and the theory of isomorphs. This is because the lines of constant ratio of the density divided by its value at the freezing point (at the same temperature), $\rho/\rho_{\rm fr}$, on the LJ system phase diagram are approximately characterized by constant values of excess entropy.    

Then, we have observed that the freezing density scaling holds for liquid argon, krypton, and xenon. The reduced viscosity and thermal conductivity coefficients along different isotherms are coinciding when plotted as functions of $\rho/\rho_{\rm fr}$ for each of the considered liquids. Freezing density scaling of the LJ type lies reasonably close to the various reference data, especially taking into account they uncertainty.   

Specific heat at constant volume also exhibits a quasi-universal behaviour in the dense fluid regime at $\rho/\rho_{\rm fr}\gtrsim 0.6$ ($s^*\lesssim -2$). A simple empirical fit is proposed which provides a unified description of reference data.  

Quasi-universality of the reduced viscosity and thermal conductivity coefficients at their minima and freezing density is discussed. The freezing point value of the reduced thermal conductivity coefficient is in excellent agreement with the prediction of the vibrational model of heat transfer. 

These results provide a consistent picture on the transport properties of simple monatomic fluids with steep isotropic pairwise interactions and represent an important step toward better understanding transport of various substances across their phase diagrams.         

The authors declare no conflict of interests.

Data sharing is not applicable to this article as no new data were created or analyzed in this study. 




\bibliography{SE_Ref}

\end{document}